# Band Topology, Orbital Phase Winding, and Selection Rules in Excitonic Physics in Two Dimensions

Ting Cao[+], Meng Wu[+], and Steven G. Louie[*]

*Department of Physics, University of California at Berkeley, Berkeley, California 94720, USA, and Materials Sciences Division, Lawrence Berkeley National Laboratory, Berkeley, California 94720, USA.*

[*] sglouie@berkeley.edu

**Abstract**:

We show that band topology can dramatically change the photophysics of two-dimensional (2D) semiconductors. For systems in which states near the band extrema are of multiple orbitals character and the spinors describing the orbital components (pseudospins) pick up nonzero winding numbers (topological invariants) around the extremal k-point, the optical strength and nature (i.e., helicity) of the excitonic states are dictated by the *optical matrix element* winding number, a unique and heretofore unrecognized characteristic. We illustrate these findings in three gapped graphene systems – monolayer graphene with inequivalent sublattices and biased bi- and tri-layer graphene, where the pseudospin textures manifest into a unique optical matrix element winding pattern associated with different valley and photon circular polarization. This winding-number physics leads to novel exciton series and optical selection rules, with each valley hosting multiple bright excitons coupled to light of different helicity. This valley-exciton selective circular dichroism can be unambiguously detected using optical spectroscopy.

An exciton in a semiconductor is an excited state with an electron-hole pair bound by their mutual Coulomb interaction [1]. Owing to the similarity between the electron-hole binding in a semiconductor and electron-proton binding in a hydrogen atom, the hydrogenic model and their variants (for example, including electron-hole-separation dependent screening effects) are usually adopted in describing excitons in various dimensions, when the electron-hole correlation length of the exciton of interest is large compared to the unit cell size. Within this picture, the envelope functions of the excitonic states are hydrogen-like wavefunctions with even or odd parity and characterized by a series of quantum numbers. In linear optical spectroscopy, an exciton may be created or annihilated by absorbing or emitting a photon, respectively. Such coupling is allowed if the full many-body excitonic states have different parity from the ground state (these states are called optically active or bright excitons). For conventional semiconductors in which the electron (hole) states forming the exciton are in a conduction (valence) band of single orbital character, this parity law together with the hydrogenic picture leads to the well-known conventional optical selection rules: in dipole-allowed materials (e.g., GaAs, monolayer transition metal dichalcogenide, etc.), *s*-like excitons are optically active, whereas *p*-like excitons are optically inactive [1-4]. In dipole-forbidden materials (e.g., $Cu_2O$), the optically active excitons are *p*-like states, while *s*-like states are optically inactive [5].

However, for many reduced-dimensional systems of current interest, the states near the band extrema are of multiple orbitals character, and the bands can have nontrivial topological characteristics. Such nontrivial topological bands may be characterized by the behavior of the amplitudes of the orbitals that compose a band state, viewed as a multi-component spinor (the pseudospin) in **k**-space. The orbital pseudospins of the electron and hole states can develop a complex texture with respect to the crystal momentum (***k***) around the band extrema [6-11]. The pseudospin texture (viewed as a spinor field of ***k***) could in principle affect the energy levels, optical selection rules, and many other properties of the excitons. Recent studies have shown that Berry curvature flux leads to a fine energy-level splitting of otherwise doubly degenerate hydrogenic 2p excitons in monolayer transition metal dichalcogenides [12, 13]. Yet, it remains unexplored whether central properties such as the optical selection rules are altered in materials with topological band characteristics.

We show here that the conventional optical selection rules, referencing to the exciton envelope functions, are *not* valid for systems with nontrivial band topology; they need to be distinctly replaced, incorporating topological effects. In the important class of 2D materials in which the pseudospins of states near the band extrema gain a nonzero winding number (topological invariant) as the carrier adiabatically traverses around the extremal **k**-point (e.g., the K or K' valley in gapped graphene systems), a highly unconventional exciton series appears and exhibits novel valley-dependent optical selection rules and other photo-activities.

The exciton energies and wavefunctions in a semiconductor may be obtained from the solutions of the Bethe-Salpeter equation (BSE) of the interacting two-particle Green's function [14]:

$$A_{\bm{k}}^S(E_{c,\bm{k}} - E_{v,\bm{k}}) + \Sigma_{\bm{k}'} A_{\bm{k}'}^S \langle cv, \bm{k} | \widehat{K}^{eh} | cv, \bm{k}' \rangle = A_{\bm{k}} \Omega^S, \qquad (1)$$

where $E_{c,\bm{k}}$ and $E_{v,\bm{k}}$ are quasiparticle energies of an electron in the conduction band and negative of the quasiparticle energy of a hole in the valence band, $A_{\bm{k}}$ describes the **k**-space exciton envelope function, and $|cv, \bm{k}\rangle$ corresponds to a free electron-hole pair (a non-interacting inter-band transition state) at the point $\bm{k}$ in the Brillouin zone (BZ). $\widehat{K}^{eh}$ is the electron-hole interaction kernel, containing a direct electron-hole attractive screened Coulomb term and a repulsive exchange bare Coulomb term. $\Omega^S$ is the excitation energy of the exciton eigenstate $|S\rangle$. For notational simplicity, we only consider here a single conduction and a single valence band. Generalization to the multiband case is straightforward, and our explicit *ab initio* results given below were performed with multiple valence and conduction bands.

The eigenstate of exciton $S$ is a coherent superposition of free electron-hole pairs at different **k** points, and is denoted by $|S\rangle = \Sigma_{\bm{k}} A_{\bm{k}}^S |cv, \bm{k}\rangle$. The oscillator strength that relates to the intensity for optical transition to exciton $S$ is given by,

$$I_e^S = \frac{2\left|\Sigma_k A_k^S \mathbf{e} \cdot \langle \phi_{c,k}|\hat{\mathbf{p}}|\phi_{v,k}\rangle\right|^2}{\Omega^S}, \tag{2}$$

where $\mathbf{e}$ is the photon polarization unit vector, and $\langle \phi_{c,k}|\hat{\mathbf{p}}|\phi_{v,k}\rangle$ the inter-band optical matrix element between the conduction band state $|\phi_{c,k}\rangle$ and valence band state $|\phi_{v,k}\rangle$.

Although the exciton energies and oscillator strengths are physical observables and thus gauge-invariant, the individual components in Eq. 2 (the exciton envelope functions in **k**-space and the inter-band optical matrix elements) may look different depending on a chosen gauge. This arises because $|cv, \mathbf{k}\rangle$ could have an arbitrary phase, which would be canceled out by the complex conjugate of the same phase in $A_k$. The gauge arbitrariness can be eliminated by requiring $A_k$ of the lowest-energy *s*-like excitonic state to be that of a hydrogen-like *s* orbital. Under this well-defined and smooth gauge, we find that an analysis of Eq. 2 illuminates clearly the physical role of the exciton envelope function and of the topological characteristics of the inter-band optical matrix elements in optical transitions. In dipole-allowed conventional semiconductors, the inter-band optical matrix elements are nearly a constant around the extremal **k**-point [1]. Therefore, only *s*-like excitons have non-zero oscillator strength, as its envelope function in **k**-space is isotropic in phase (i.e., no phase winding around the extremal **k**-point).

Having topologically nontrivial bands in 2D with associated pseudospin texture of nonzero winding numbers will lead to both magnitude and phase modulations of the inter-band optical matrix elements with $\mathbf{k}$, represented by a 2D vector field with a certain winding pattern. To illustrate this effect, we decompose the inter-band optical matrix element $\langle \phi_{c,k}|\hat{\mathbf{p}}|\phi_{v,k}\rangle$ into the two irreducible cylindrical components, $p_{k+} = \mathbf{e}_+ \cdot \langle \phi_{c,k}|\hat{\mathbf{p}}|\phi_{v,k}\rangle$ and $p_{k-} = \mathbf{e}_- \cdot \langle \phi_{c,k}|\hat{\mathbf{p}}|\phi_{v,k}\rangle$, which correspond to coupling to left- and right-circularly polarized photon modes ($\sigma_-$ and $\sigma_+$), respectively. For topologically nontrivial bands, as illustrated below, $p_{k+}$ and $p_{k-}$ are typically non-zero (except at the extremal **k**-point), and can be viewed as two vector fields that may differ in their winding patterns. (We note that the inter-band optical matrix elements $p_{k\pm}$ are complex quantities determined by the bands and independent of the specific excitonic states.)

We shall show that the brightness and helicity of an exciton are dictated by the phase winding of the exciton envelope function and that of the inter-band optical matrix elements. For an excitonic state of which the **k**-space envelope function $A_k$ is a highly localized function (Wannier excitons) around an extremal **k**-point, $A_k$ and $p_{k\pm}$ in the relevant small part of the BZ are dominated by a cylindrical angular phase dependence of $\sim e^{im\theta_k}$ and $e^{il_\pm \theta_k}$, respectively ($\theta_k$ is the angle of **k** defined with respect to the *x*-axis). Here, and in subsequent discussion, we shall define **k** as the wavevector measured from the extremal **k**-point. Thus, $m$ is cylindrical angular quantum number of the exciton envelope function and $l_\pm$ are the winding number of $p_{k\pm}$. Following Eq. 2, the oscillator strength for an optical transition to an excitonic state $S$ by $\sigma_\pm$ photon is $I^S_{\sigma_\pm} = \frac{2\left|\Sigma_k f(|\mathbf{k}|)e^{i(m+l_\mp)\theta_k}\right|^2}{\Omega^S}$, where $f(|\mathbf{k}|)$ is the radial part in the summation. $I^S_{\sigma_\pm}$ is thus non-zero *only* when

$$m = -l_\mp. \tag{3}$$

This set of selection rules is thus distinctly different from conventional semiconductors. For a system with discrete n-fold rotational symmetry, the general selection rule is: $m + l_\mp = 0 \bmod n$. (A generalization to systems with discrete rotational symmetries is given in the Supplementary Information Section I.) As a result, excitons with different angular quantum numbers (i.e., different $m$) would couple differently to $p_{k+}$ and $p_{k-}$, causing multiple bright excitons each accessible by $\sigma_-$ or $\sigma_+$ photons.

An ideal set of materials to illustrate the predicted novel excitonic physics is the gapped graphene systems, in which a bandgap and a layer-number-dependent pseudospin texture emerge from an induced broken inversion symmetry that may be tuned. We consider three (already experimentally achieved) systems based on 1 to 3 layers of graphene [15-18]. For monolayer graphene, the inversion symmetry is broken by placing the graphene layer on top of a monolayer of hexagonal boron nitride, with the two sublattices A and B of graphene sitting directly on top of the boron and nitrogen atoms, respectively. For bilayer (in a Bernal stacking order) and trilayer graphene (in a rhombohedral stacking order), inversion symmetry is broken by applying an external electric field along the out-of-plane direction. In our *ab initio* GW-BSE calculations presented below, the applied electric field was set to 0.13 eV/Å, an experimentally studied value. Modifying the applied electric field strength does not change the physics discussed here.

For the gapped graphene systems studied, density functional theory (DFT) calculations are performed within the local density approximation (LDA) formalism using the Quantum ESPRESSO package [19] to determine their ground-state properties. First-principles GW [20] and GW-BSE [14] methods are employed to calculate the quasiparticle band structure and excitonic states, respectively, using the BerkeleyGW package [21]. The dielectric matrix for the screened Coulomb interaction is constructed with a 2D truncation scheme [22] and with an energy cutoff of 8 Ry. Close scrutiny is needed for the BZ sampling in the excited-state calculations. For calculations of the quasiparticle band structure, a $150 \times 150$ **k**-point mesh in the BZ is necessary to converge the bandgap within 3 meV. For the calculation of excitons, a patched sampling scheme is used to solve the BSE for the excitonic states in the individual K and K' valleys. The sampling density is equivalent to a uniform $450 \times 450$ **k**-point mesh in the BZ. For monolayer graphene, a $450 \times 450$ **k**-point mesh is moreover interpolated into a $1500 \times 1500$ mesh to converge the exciton energy levels to within 2 meV.

The gapped graphene systems of 1, 2, and 3 atomic layers studied have GW quasiparticle bandgaps of ~ 130 meV, 159 meV, and 185 meV [Fig. 1(a-c)], respectively. These values are much larger than their corresponding DFT-LDA Kohn-Sham bandgaps of ~ 62 meV, 90 meV and 118 meV, respectively, owing to electron self-energy effects. For bilayer and trilayer graphene, the top valence and bottom conduction bands at the K and K' valleys develop a Mexican-hat-like shape. The pseudospin texture of the states in bilayer graphene is schematically shown in Fig. 1d, where the amplitude of the carbon π orbitals develop a phase winding around the band extremum [23].

The very different pseudospin texture of the bands in the three gapped graphene systems gives a strong layer-number and valley-index dependent, inter-band optical matrix element winding pattern for each. We show in Fig. 2 the winding pattern of $p_{k+}$ and $p_{k-}$ in the K valley, defined using the gauge procedure as describe above. In the plot, the complex quantity $p_{k+}$ or $p_{k-}$ (which is given by a magnitude and a phase $\phi_k$) are represented by an arrow with its length proportional to the magnitude and its orientation pointing along the direction with angle $\phi_k$ to the *x*-axis. In monolayer graphene with inequivalent A/B sublattices [Fig. 2a, b], $p_{k+}$ is nearly constant in magnitude and phase (arrows with constant length and orientation) and has a winding number = 0 for any contours enclosing K, whereas $p_{k-}$ is much smaller in magnitude and its phase (the orientation of the arrows) winds clockwise around the K point twice (winding number = -2) after completing any contour enclosing K. This analysis, making use of the selection rules deduced above, predicts an optically active *s* exciton series, as well as a weakly active *d* exciton series. In biased bilayer graphene, the pseudospin texture [Fig. 1d] leads to a winding number = 1 for the inter-band optical matrix element $p_{k-}$ [Fig. 2e]. Compared with $p_{k-}$, $p_{k+}$ [Fig. 2d] is much smaller in magnitude, but remains constant in both magnitude and phase around the K point (winding number = 0). It therefore predicts: (i) unlike the case of monolayer gapped graphene, the *p* exciton series is now optically very active; (ii) the *s* exciton series are still somewhat optically active, but having a much smaller oscillator strength than the *p* exciton series; and (iii) importantly, the *s* excitons and *p* excitons at a given valley (K or K') have opposite helicity in biased bilayer graphene. The inter-band optical matrix elements in biased trilayer graphene have even more features [Fig. 2g, h], leading to a winding number of 1 and 2 for $p_{k+}$ and $p_{k-}$, respectively, at the K valley. (Details in Supplementary Information Section II.) The GW-BSE 1*s* exciton envelope functions of the three gapped graphene systems studied are shown in Fig. 2c, f, and i. Our new selection-rule predictions based on topological effects are completely borne out by our explicit *ab initio* GW-BSE calculations of the optical absorption spectra.

The physics of inter-band optical matrix element winding thus leads to novel exciton series in the gapped graphene systems, with each valley hosting multiple optically active excitons having different helicity. We show in Fig. 3 the *ab initio* GW-BSE calculated energy level, helicity, and oscillator strength of the first six lowest-energy excitons in the K- and K'-valley of each system. The calculated binding energies of the lowest exciton state of the 1-, 2-, and 3-layer systems are 34 meV, 52 meV, and 45 meV, respectively. In monolayer graphene with inequivalent sublattices [Fig. 3a], as expected, the *s*-like excitons are optically bright. The 1*s* exciton in the K and K' valleys can be selectively excited by $\sigma_-$ and $\sigma_+$ light, respectively, similar to monolayer transition metal dichalcogenides [24-27]. In biased bilayer graphene [Fig. 3b], however, the optically most active exciton becomes a 2*p* state that is located at 13 meV above the lowest energy 1*s* state, with an oscillator strength ~ 20 times larger than that of the 1*s* exciton. Moreover, the helicity of the 2*p* state is opposite to that of the 1*s* state, a feature that is predicted from the inter-band optical matrix element winding patterns depicted in Fig. 2d and Fig. 2e. In the biased trilayer graphene [Fig. 3c], the lowest energy 1*s* exciton is optically inactive from the matrix element winding patterns in Figs. 2g and 2h. Due to a significant deviation of the band dispersion from a parabola, we are no longer able to associate the higher energy excited excitonic states with a clear principal quantum number. However, a pair of nearly degenerate excitons with *p*-like and *d*-like orbital characters could still be identified, located at ~ 9 meV above the 1*s* state. They are $\sigma_+$ polarized, and couple strongly (optically bright) to the ground state via $p_{k-}$ in Fig. 2h directly, or through a trigonal warping effect. (There is also a weakly active *p*-like exciton at ~ 4 meV above the 1*s* state. Details in Supplementary Information Section II.) In all three cases, the helicity of every bright exciton in the K'-valley is opposite to that of a degenerate-in-energy partner exciton in the K-valley due to time reversal symmetry.

We now show how our findings on the predicted novel 2D excitonic physics may be experimentally verified by polarization-resolved optical spectroscopy. As phonon-assisted intravalley exciton energy relaxation is much more efficient than phonon-assisted intervalley exciton energy relaxation [25, 28, 29], optically created excitons in one valley will predominantly relax to the lowest energy exciton in the same valley. Taking biased bilayer graphene as an example, resonant $\sigma_-$ excitations of the K-valley 1*s* exciton will induce a $\sigma_-$ photoluminescence from the excited excitons themselves, whereas resonant $\sigma_-$ excitations of the K'-valley 2*p* exciton will induce photoluminescence from the K'-valley 1*s* exciton following energy relaxation from the 2*p* state to the 1*s* state. As the helicity of the 1*s* exciton is opposite to that of the 2*p* exciton in the same valley [Fig. 3b], the latter excitations would produce a $\sigma_+$ photoluminescence. This is a predicted new phenomenon in biased bilayer graphene that is quite different from the behavior of photoluminescence in monolayer transition metal dichalcogenides or gapped graphene [24-27], because the helicity of the luminescence light for the former would depend not only on the polarization of the incident light, but also on the excitation energy (whether it is to the $1s$ or $2p$ exciton energy).

In summary, we have presented results of novel exciton series and optical selection rules arising from band topological effects in 2D semiconductors. Our work reveals another important manifestation of band topology in the physical properties of materials; it also open opportunities for use of these effects in gapped graphene systems for potential valleytronic applications.

**Acknowledgments:**

This work was primarily supported by the van der Waals Heterostructures Program at the Lawrence Berkeley National Lab through the Office of Science, Office of Basic Energy Sciences, U.S. Department of Energy under Contract No. DE-AC02-05CH11231. Supported by the National Science Foundation under Grant No. DMR-1508412 is also acknowledged, which provided for the *GW* calculations. Computational resources were provided by the DOE at Lawrence Berkeley National Laboratory's NERSC facility and the NSF through XSEDE resources at NICS. T. C. thanks L. Ju for helpful discussions.

[+] T. C. and M. W. contribute equally to this work.

**Figures**

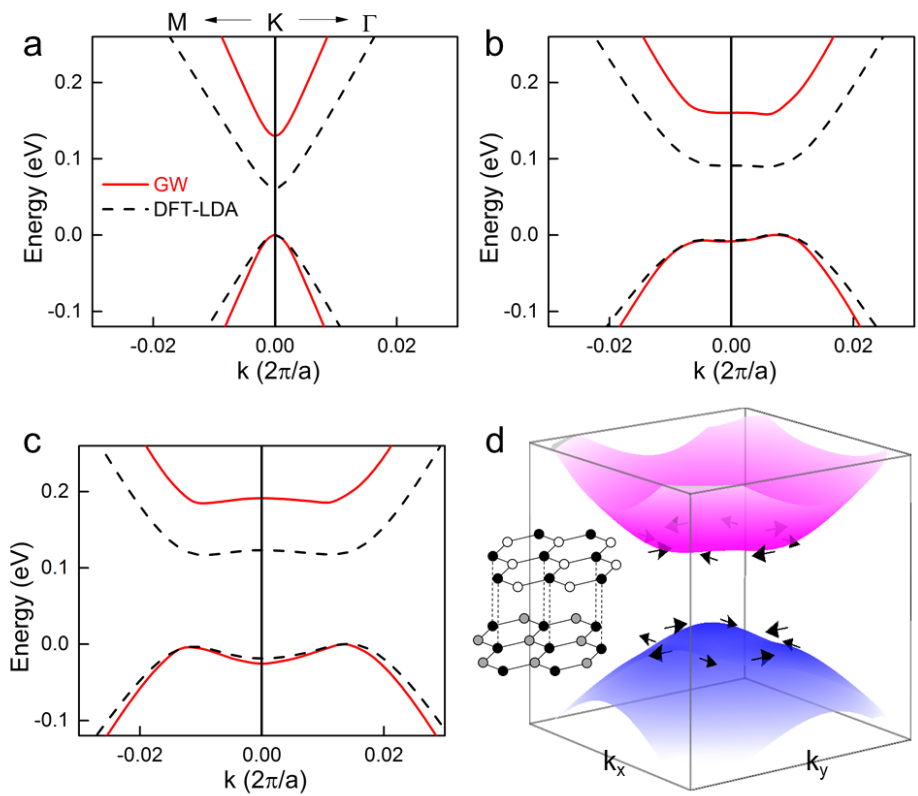

FIG. 1. Calculated band structure and orbital phase winding of gapped graphene systems. Bottom conduction band and top valence band of monolayer graphene with broken A/B sublattice symmetry (a), biased Bernal-stacked bilayer graphene (b), and biased rhombohedral-stacked trilayer graphene (c). Red solid lines and black dashed lines are GW and DFT-LDA bands, respectively. The K point is set at k = 0. Positive and negative k values denote the K-Γ and K-M direction, respectively. (d) Orbital pseudospin phase winding in biased bilayer graphene. Inset: structure of biased bilayer graphene. The carbon atoms forming bonds with a neighboring layer are colored black.

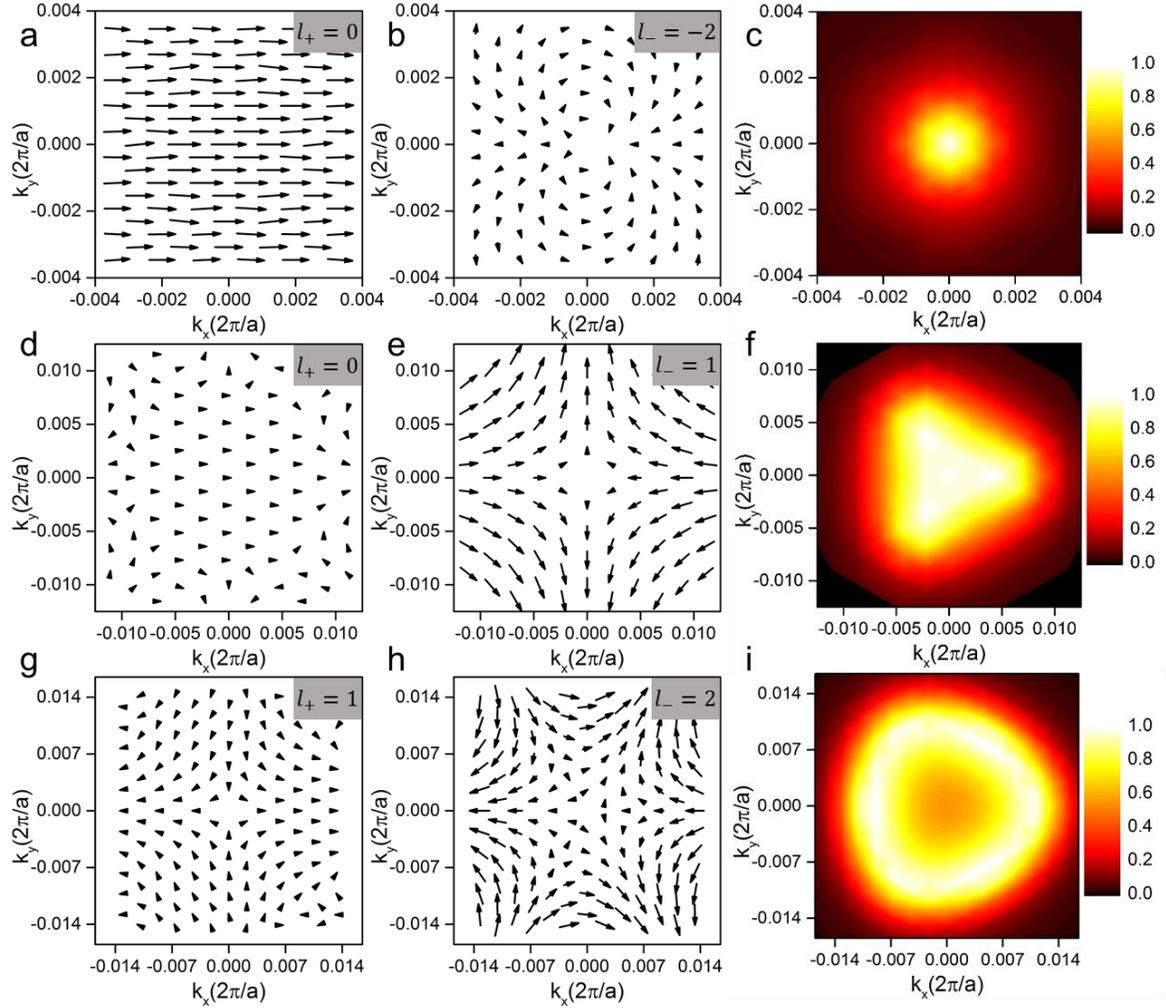

FIG. 2. K-valley inter-band optical transition matrix elements and 1$s$ exciton envelope function in **k**-space. The K point is placed at the origin. Optical inter-band transition matrix element and its winding number for light of (a) left circular polarization $p_{\mathbf{k}+}$ and (b) right circular polarization $p_{\mathbf{k}-}$ in monolayer graphene with inequivalent sublattices. The direction and length of the arrow denote respectively the phase and the magnitude of the matrix element. (d) $p_{\mathbf{k}+}$ and (e) $p_{\mathbf{k}-}$ in biased bilayer graphene. (g) $p_{\mathbf{k}+}$ and (h) $p_{\mathbf{k}-}$ in biased trilayer graphene. (c, f, i) 1$s$ exciton envelope function in **k**-space in gapped monolayer graphene, biased bilayer graphene, and biased trilayer graphene, respectively. The envelope functions show the magnitude of the free electron-hole pair excitation at each **k**, normalized to its largest value in each plot.

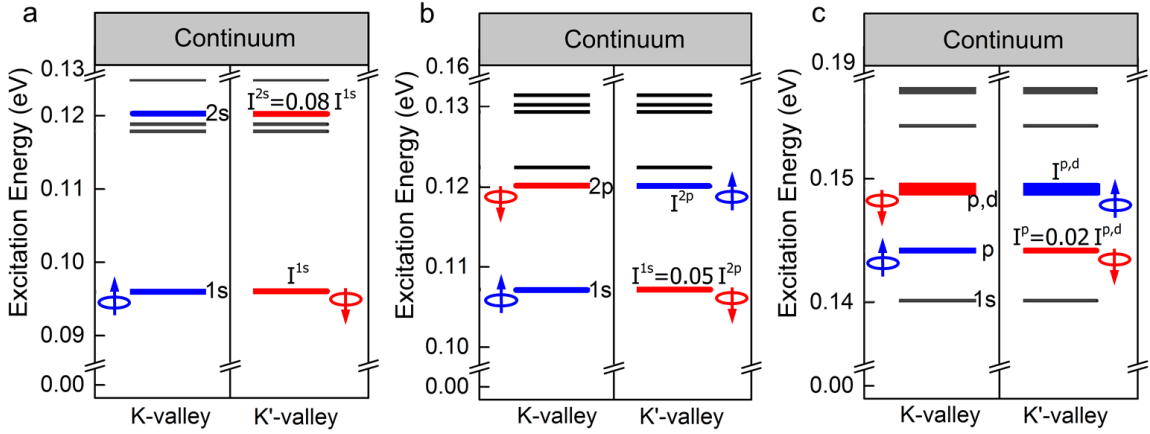

FIG. 3. K-valley and K'-valley exciton energy levels and valley-exciton selective circular dichroism in (a) monolayer graphene with inequivalent A/B sublattices, (b) biased bilayer graphene, and (c) biased trilayer graphene. Left and right part of each panel depicts the K-valley and K'-valley exciton energy levels, respectively. The first six lowest-energy excitons are shown in each plot. Black lines indicate dark states (with maximum oscillator strength < 1% of the brightest exciton in each plot). The oscillator strength I of each bright state is expressed in terms of that of the brightest state, for unpolarized light. Blue and red lines (or circles) indicate bright states with left- and right-helicity, respectively.